\documentclass[twocolumn,amsmath,amssymb]{snp}
\usepackage{graphicx}% Include figure files
\usepackage{dcolumn}% Align table columns on decimal point
\usepackage{bm}% bold math
\begin{document}

\title{\Large Dynamical Study of Multifragmentation and Related Phenomena in Heavy-Ion Collisions} %line breaks with \\

\author{\large Yogesh K. Vermani
\footnote{Present Address: ITM University, Gurgaon-122017, INDIA}}
 \email{yugs80@gmail.com}

\affiliation{Department of Physics, Panjab University,
Chandigarh-160014, INDIA}

\maketitle

\section{Introduction \& Methodology}
Study of heavy-ion collisions at intermediate energies has now
become important tool to investigate reaction mechanism behind
collective expansion and origin of fragments. Apart from this, it
also becomes possible to infer nuclear matter equation of state
(EoS) \cite{dan79,shi, dan93}.

For the present thesis work, we shall employ \emph{quantum
molecular dynamics} (QMD) model \cite{aich} to simulate the
nucleus-nucleus collisions. This model is well suited to study
A-particles system where nucleon-nucleon correlations become
important during the collision process. The QMD approach treats
the nucleons as gaussian wave packets with total nuclear wave
function given as:
\begin{eqnarray}
\Phi &=&{\prod_{i=1}^{A}}{\psi}_i({\bf r},{\bf r}_i,{\bf p}_i,t), \nonumber \\
     &=&{\prod_{i=1}^{A}}\frac{1}{(2\pi L)^{3/4}} e^{-({\bf r}-{\bf r}_i(t))^2/4L} \cdot e^{\frac{\iota}{\hbar}{\bf p}_i(t)\cdot{\bf r}}.
\end{eqnarray}
Note that antisymmetrization is neglected here. The width of
gaussian wave packet is taken to be independent of time with value
$L=1.08 ~fm^{2}$. The total Hamiltonian of the A-particles system
is given as:
\begin{equation}
\langle H \rangle =\sum^{A}_{i=1}{\frac{{\bf p}_i^2}{2m_i}} +
V^{Sk} + V^{Col} +V^{Yuk} \label{H}
\end{equation}
The interaction potential in Eq.(\ref{H}) consists of density
dependent Skyrme interaction supplemented with Coulomb and Yukawa
potentials.
\begin{figure} [!b]
\centering \setlength{\abovecaptionskip}{-1.0cm}
\setlength{\belowcaptionskip}{0.5cm} \vskip -1.0cm
\includegraphics [width=72mm]{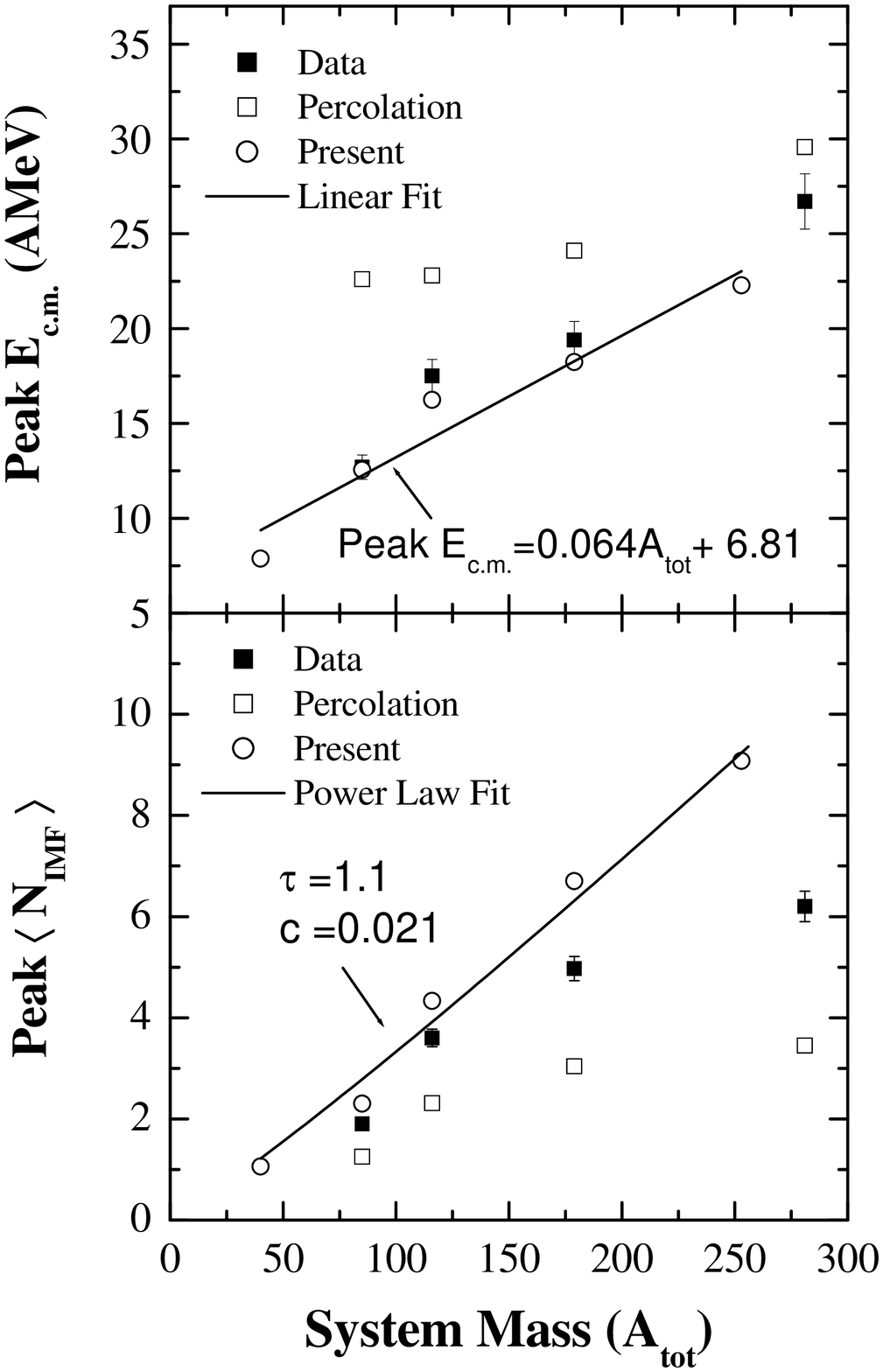}% Here is how to import EPS art
\vskip 0.9cm \caption {Peak $E_{c.m.}$ and peak $\langle N_{IMF}
\rangle$ as a function of total system mass $A_{tot}$. Open and
solid squares depict the percolation calculations and experimental
data points, respectively \cite{sis}.}
\end{figure}

\section{Results and discussions}
In the first part of thesis, we shall deal with fragment emission
in central collisions studied as a function of beam energy and
system mass. Central collisions are also important candidate in
view of exploring collective expansion and squeeze out phenomena
\cite{dan93}. We have simulated the central collisions of
$^{20}Ne+^{20}Ne$, $^{40}Ar+^{45}Sc$, $^{58}Ni+^{58}Ni$,
$^{86}Kr+^{93}Nb$, $^{129}Xe+^{124}Sn$, and $^{197}Au+^{197}Au$.
Our model calculations for the multiplicity of intermediate mass
fragments (IMFs) as a function of beam energy available in the
center-of-mass frame agree with the experimental trends observed
on MSU 4$\pi$-array set-up. We further plot the peak $E_{c.m.}$
(at which maximal emission occurs) and peak IMF multiplicity as a
function of total system mass $A_{tot}$ as shown in Fig.1.
Interestingly, peak IMF multiplicity is observed to follow power
law of form: $c A^{\tau}_{tot}$, with exponent $\tau$ close to
unity \cite{yugs}.
\begin{figure} [!t]
\centering \setlength{\abovecaptionskip}{-1.0cm}
\setlength{\belowcaptionskip}{0.5cm} \vskip -1.0cm
\includegraphics [width=72mm]{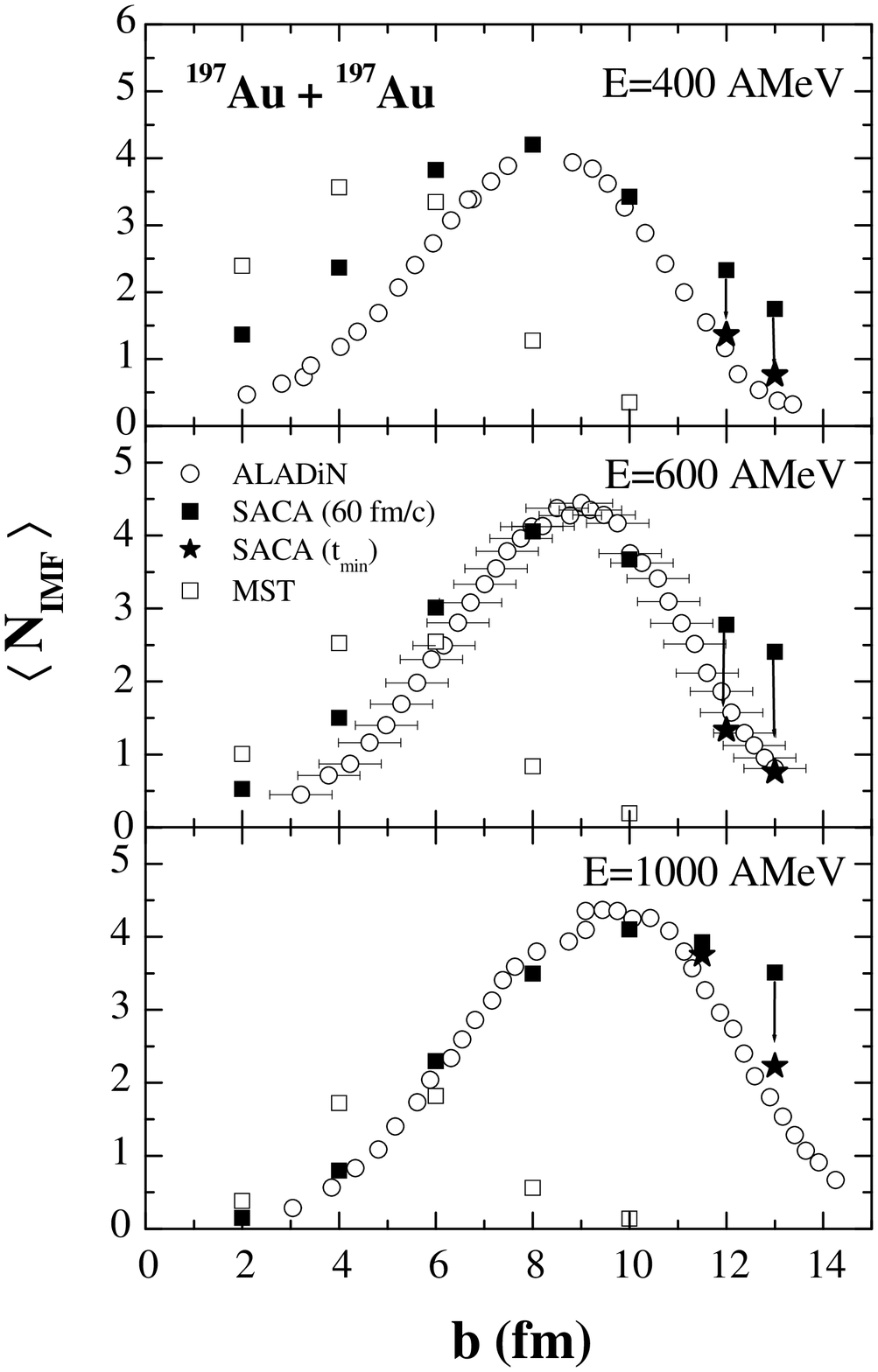}% Here is how to import EPS art
\vskip 0.5cm \caption {The mean IMF multiplicity $\langle N_{IMF}
\rangle$ vs impact parameter b for the reaction of
$^{197}Au+^{197}Au$ at 400, 600 and 1000 AMeV, respectively.}
\vskip -1.0cm
\end{figure}

Next we try to understand the clusterization mechanism in
spectator matter fragmentation using \emph{simulated annealing
clusterization algorithm} (SACA) advanced by Puri \emph{et al}
\cite{jcp,epl}. In this approach, pre-clusters obtained with
\emph{minimum spanning tree} (MST) method are subjected to a
binding energy condition \cite{jcp,epl}:
\begin{eqnarray}
\zeta_{a}=\frac{1}{N_{f}}\sum_{i=1}^{N_{f}}\left[\sqrt{\left(\textbf{p}_{i}-\textbf{P}_{N_{f}}^{cm}\right)^{2}+m_{i}^{2}}-m_{i}
+ \right . \nonumber \\
\left. \frac{1}{2}\sum_{j\neq i}^{N_{f}}V_{ij}
\left(\textbf{r}_{i},\textbf{r}_{j}\right)\right]< -E_{bind},
\label{be}
\end{eqnarray}
with $E_{bind}$ = 4.0 MeV if $N_{f}\geq3$, else $E_{bind} = 0$. In
this equation, $E_{bind}$ is the fragment's binding energy per
nucleon, $N_{f}$ is the number of nucleons in a fragment, and
$P_{N_{f}}^{cm}$ is the center-of-mass momentum of the fragment.
Using this approach, we study the spectator matter fragmentation
in peripheral $^{197}Au+^{197}Au$ collisions as a function of
impact parameter. The IMFs yields obtained using SACA method are
then compared with conventional MST algorithm and ALADiN
experimental data (See Fig.2). Remarkably, SACA calculations
explain the universality feature in IMF production in the incident
energy range 400-1000 AMeV quite well \cite{epl}. Earlier
recognition of fragments structure (around 60 fm/c) also points
towards dynamical origin of fragments. In other words, system
doesn't have enough time span to undergo complete equilibration.

We shall also highlight the importance of momentum dependent
interactions in probing nuclear EoS via intermediate energy
heavy-ion collisions. Estimation of baryonic entropy shall also be
attempted within QMD approach using composite particles yield
ratios. \\

\end{document}